# Giant electromechanical response from defective non-ferroelectric epitaxial BaTiO$_3$ integrated on Si (100)

Sandeep Vura[1]*, Shubham Kumar Parate[1]*, Subhajit Pal[1,2], Upanya Khandelwal[1], Rajeev Kumar Rai[1,3], Sri Harsha Molleti[1], Vishnu Kumar[1], Rama Satya Sandilya Ventrapragada[1], Girish Patil [1], Mudit Jain[1], Ambresh Mallya[1], Majid Ahmadi[4], Bart Kooi[4,5], Sushobhan Avasthi[1], Rajeev Ranjan[6], Srinivasan Raghavan[1] , Saurabh Chandorkar[1],  Pavan Nukala[1].

**Author Affiliations**

[1]. Center for Nano Science and Engineering, Indian Institute of Science, Bengaluru-560012, India

[2]. School of Engineering and Materials Science, Queen Mary University of London, London E1 4NS, United Kingdom

[3]. Materials Science and Engineering, University of Pennsylvania, 3231 Walnut Street, Philadelphia, 19104

[4]. Zernike Institute for Advanced Materials, University of Groningen, Groningen, 9747AG, The Netherlands

[5]. CogniGron center, University of Groningen, Groningen, 9747 AG, The Netherlands.

*indicates equal contribution

[6]. Materials Engineering, Indian Institute of Science, Bengaluru, 560012- India

Corresponding Author Address

Pavan Nukala, Center for Nano Science and Engineering, Indian Institute of Science, Bangalore-560012, India, email: pnukala@iisc.ac.in

Sandeep Vura, Center for Nano Science and Engineering, Indian Institute of Science, Bangalore-560012, India, email: sandeepv@iisc.ac.in

**Abstract:**

Lead-free, silicon compatible materials showing large electromechanical responses comparable to, or better than conventional relaxor ferroelectrics, are desirable for various nanoelectromechanical devices and applications. Defect-engineered electrostriction has recently been gaining popularity to obtain enhanced electromechanical responses at sub 100 Hz frequencies. Here, we report record values of electrostrictive strain coefficients ($M_{31}$) at frequencies as large as 5 kHz ($1.04\times10^{-14}$ $m^2/V^2$ at 1 kHz, and $3.87\times10^{-15}$ $m^2/V^2$ at 5 kHz) using A-site and oxygen-deficient barium titanate thin-films, epitaxially integrated onto Si. The effect is robust and retained even after cycling the devices >5000 times. Our perovskite films are non-ferroelectric, exhibit a different symmetry compared to stoichiometric $BaTiO_3$ and are characterized by twin boundaries and nano polar-like regions. We show that the dielectric relaxation arising from the defect-induced features correlates very well with the observed giant electrostrictive response. These films show large coefficient of thermal expansion ($2.36 \times 10^{-5}$/K), which along with the giant $M_{31}$ implies a considerable increase in the lattice anharmonicity induced by the defects. Our work provides a crucial step forward towards formulating guidelines to engineer large electromechanical responses even at higher frequencies in lead-free thin films.

**Introduction**

The quest for lead-free piezoelectric materials with a large electromechanical (EM) response is very important for nano electromechanical systems (NEMS) devices such as actuators, ultrasonic transducers, sensors, energy harvesters, nanopositioners, and micro-robotics[1–4]. A common strategy to increase piezoelectric response from non-centrosymmetric materials is to flatten the thermodynamic energy profile easing polarization rotation [5–8]. Such an energy landscape is seen at the morphotropic phase boundary (MPB) in materials such as $Pb(Zr,Ti)O_3$ (PZT) and relaxor ferroelectrics, and in materials with many coexisting local orders ($K_{0.5}Na_{0.5}NbO_3$-$xBaTiO_3$ (KNN-BT, for e.g.)[5–10]. In classic ferroelectrics, domain wall motion is a major contributor to the EM response, especially at lower frequencies[11,12]. In centrosymmetric systems, especially thin films, field-induced piezoelectricity has been engineered via various strategies such as creating asymmetric contacts[13], multiple interfaces[14–16] and defects[15–18].

Recently, defect-based strategies are gaining tremendous traction to engineer materials with large non-classical EM responses, especially at low frequencies (< 100 Hz). "Giant" electrostriction[4] has been first reported[19] in 20% Gd doped $CeO_2$ in 2012. It has by now been established that the response of electroactive defect complexes in non-dilute concentrations to an external electric field, and their corresponding elastic dipoles results in substantial electrostrain[14,17]. Such an effect, which is second order in nature or electrostrictive, has also been engineered in other defective oxide systems such as Nb and Y stabilized $Bi_2O_3$[20] $La_2Mo_2O_6$[21] and so on[4]. Electromechanical effects (both first and second order) were also observed in thin-film systems such as $BaTiO_3$ through oxygen vacancy-induced chemical expansion[18]. Colossal piezoelectric coefficients in centrosymmetric gadolinium-doped $CeO_{2-x}$ ($d_{33}$~200,000 pm/V) at ~1 mHz have been attributed to field-induced defect/ion motion at those frequencies[17]. In the same system, at slightly larger frequencies (100- 1000 Hz) polaron hopping is shown to be responsible for an effective $d_{33}$~100 pm/V, and which is comparable to response from the classic piezoelectrics such as PZT. Furthermore, enhanced field-dependent $d_{33}$ values of ~1100 pm/V were also reported in A-site deficient $NaNbO_3$ films deposited on $SrTiO_3$ replete with out-of-phase boundaries (2D defects)[15,16,22]. Defect-driven EM response dramatically reduces as frequency increases and becomes insignificant beyond 1 kHz. A unifying thread in all these studies is to engineer EM response through large concentration of electroactive defects (0D and 2D defects), which elastically interact with each other to give a coherent and large strain response.

Here, we report record second-order EM coefficients at frequencies larger than 1 kHz ($M_{31}$= 1.04 × $10^{-14}$ $m^2/V^2$ at 1 kHz, and 3.87 × $10^{-15}$ $m^2/V^2$ at 5 kHz) in heavily A-site and oxygen-deficient non-ferroelectric barium titanate ($Ba_{0.89}TiO_{3-\delta}$) thin film system. The BTO was epitaxially integrated onto Si with TiN as a buffer layer. The oxygen scavenging property of TiN buffer layer, along with suitable choice of growth parameters, allows us to obtain a very non-stoichiometric, yet stable, perovskite phase. We show that the observed giant EM response is correlated to the defect-based mechanisms giving rise to dielectric relaxation, and possibly even to electroactive twin boundary mobility. We propose that in addition to having large defect concentration, systems with significantly enhanced defect-induced dielectric constant and relaxation behavior, are good materials for large and non-classical EM response. This work is a significant step in achieving lead-free CMOS compatible materials with giant EM response.

## Results and Discussion

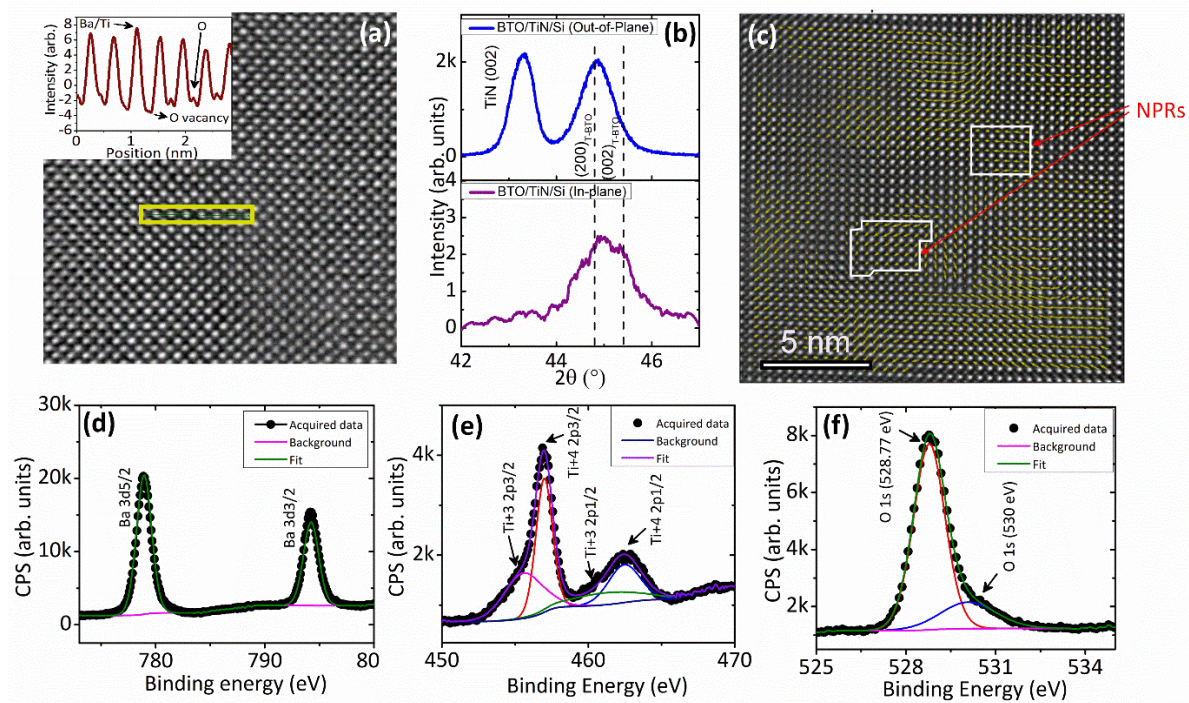

*Fig. 1: **Structural and chemical characterization of defective BTO thin films** (a) iDPC STEM image and corresponding line intensity profile from the highlighted region. Low intense oxygen peaks show up in some columns and are absent in some revealing the presence of oxygen vacancies. (b) X-ray diffraction θ-2θ scans showing the out-of-plane Bragg peaks (top) and in-plane scans (bottom). (c) High resolution HAADF STEM image of BTO overlayed with polarization map nano polar-like regions (NPRs) are enclosed in white boxes. High resolution XPS spectra with fits of (d) Ba 3d (e) Ti 2p and (f) O 1s.*

The effect of deposition conditions on epitaxy and growth of the defective complex oxides on Si studied here was previously developed by Vura et al. and has been reported elsewhere[23,24]. A buffer layer of epitaxial TiN ((100) oriented, 40-60 nm) was deposited on n++ Si (100) using reactive pulsed laser deposition in $N_2$ (99.9999% pure) atmosphere in eclipsed off-axis configuration[23] [see Methods]. Epitaxial growth of barium titanate (BTO, <001> oriented, 175-245 nm) is enabled directly on this platform (Fig S1), using standard pulsed laser deposition. All the results reported here are on samples grown at conditions described in the methods, and post-annealed in atmosphere at 500 °C. High Angle Annular Dark Field Imaging in Scanning Transmission Electron Microscopy (HAADF-STEM) and corresponding energy dispersive spectroscopy (EDS, Fig S1) analysis reveals the presence of a TiO$_x$ (18-20 nm) layer between TiN and BTO, formed via interfacial redox reaction during annealing. This reaction also helps in rendering the BTO layer oxygen deficient. Oxygen vacancies can be identified as missing oxygen contrast in iDPC STEM images and corresponding line profiles (Fig 1a). The out-of-plane and in-plane lattice parameters, obtained from θ-2θ and in-plane XRD scans (Fig 1b) correspond to 4.038 (±0.005) Å and 4.023 (±0.005) Å, respectively. The unit cell volume of the film (65.35 Å$^3$) is larger than that of bulk ceramic BTO (a=3.99 Å, c=4.04 Å, 64.31 Å$^3$), as a result of chemical expansion due to defects[24]. The polarization maps obtained from HAADF STEM (Fig 1c) images show the presence of nano polar-like regions, similar to relaxors[25]. The non-existence of a unique polarization axis reveals that our defective BTO has a different symmetry than P4mm. Our films are also replete with 2D defects such as twin boundaries (Fig S2a). Film composition is determined to be Ba$_{0.89}$TiO$_{3-δ}$ via depth-resolved X-ray photoemission spectroscopy (XPS) (Fig 1(d-f)). Peak fitting procedures reveal the presence of Ti in both 4+ and in reduced 3+ oxidation states.

On these defective BTO thin films, we fabricated interdigitated electrodes (IDE, Fig S2b (inset), Fig. S2c), with electrode separation of 20 μm, and finger length of 80 μm. A voltage applied across two terminals majorly contributes to in-plane electric field in BTO (referred to as *direction 1*), owing to the presence of a low dielectric constant insulating TiO$_x$ interfacial layer. In this configuration, by applying a large signal AC voltage, we measured the out-of-plane displacement (*direction 3*) on various devices (~10 devices) using laser doppler vibrometer (see Methods, also see Fig S2b). Voltage was cycled between V$_{max}$ (1 to 5 V) and –V$_{max}$ at different frequencies (1 kHz to 50 kHz), for >5000 cycles on each device (see Methods), and time-averaged strain response as a function of voltage is computed [Methods]. The piezoelectric tensor component we extract from these experiments, hence, is $d^*_{31}$ and the electrostrictive tensor component is M$_{31}$ (in Voigt notation).

The EM strain ($\epsilon_i$) as a function of electric field (E$_j$) can be expanded upto second order as follows:

$$\epsilon_i = d^*_{ij}E_j + M_{ij}E_j^2$$

$$\epsilon_i = (d_{ij}^*)_0 e^{i\phi_1} E_j + (M_{ij})_0 e^{i\phi_2} E_j^2$$

where $E_j(\omega) = Re((E_j)_0 e^{i\omega t})$

$(d_{ij}^*)_0$ is the amplitude of piezoelectric coefficient and $(M_{ij})_0$ is the amplitude of electrostrictive coefficient of defective BTO, $\phi_1$ and $\phi_2$ represent the phase difference between voltage (V) and piezoelectric strain ($\epsilon_i^{(1)}$); and $V^2$ and second order response ($\epsilon_i^{(2)}$), respectively.

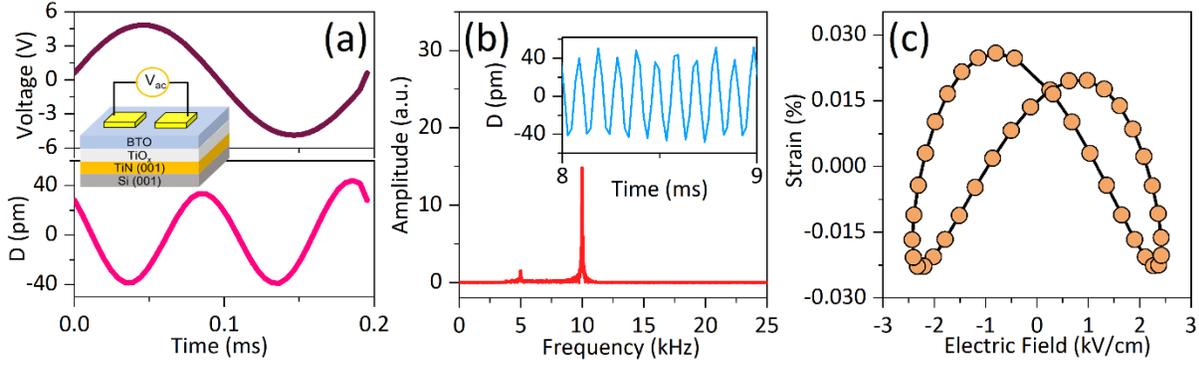

*Fig 2: **Electromechanical response of defective BTO lateral devices** (a) Averaged voltage (top) and corresponding averaged displacement (D) response (bottom) as a function of time (inset) lateral measurement setup schematic (inset). Optical micrograph of IDE and corresponding dimensions (in μm) are shown in Fig S2(b (inset) and c) (b) Fast Fourier Transform of raw displacement-time response shown in inset (c) Averaged strain-electric field response.*

The first order (harmonic) strain response is given as follows:

$$\epsilon_i^{(1)} = (d_{ij}^*)_0 e^{i(\omega t + \phi_1)} E_j$$

The second order strain response is given as follows:

$$\epsilon_i^{(2)} = (M_{ij})_0 e^{i\phi_2} (E_j)_0^2 ((\cos(\omega t))^2 = (M_{ij})_0 e^{i\phi_2} (E_j)_0^2 ((\cos(2\omega t) + 1)/2)$$

which can be further split into second harmonic component and a DC component as follows

$$\epsilon_i^{(2)}(2\omega) = \frac{Re\left((M_{ij})_0 (E_j)_0^2 e^{i(2\omega t + \phi_2)}\right)}{2}$$

$$\epsilon_i^{(2)}(DC) = \frac{Re\left((M_{ij})_0 (E_j)_0^2 e^{(i\phi_2)}\right)}{2}$$

Averaged input voltage and corresponding averaged displacement response as a function of time on lateral devices are shown in Fig 2a. The schematic representation of the device is also displayed in the inset of Fig 2a. The Fourier transform of displacement-time response (Fig 2b (inset)) shown in Fig 2b predominantly shows a second order response (and a weak first order response). This is also reflected in the butterfly-like strain-voltage plots in Fig 2c.

Fourier filtered second order strain response as a function of voltage is shown in Fig 3a. The amplitude of effective electrostrictive coefficient ($|M_{31}|$) on a representative device, calculated from Fig 3a as a function of frequency at 3 V and 5 V is shown in Fig 3b. We estimate that $|M_{31}|$ at 1 kHz and $V_{max}$=5 V is $1.04 \times 10^{-14}$ m$^2$/V$^2$, and it reduces by more than half at 5 kHz to $3.87 \times 10^{-15}$ m$^2$/V$^2$. These are record values observed at frequencies >1 kHz, as can be seen from the comparison of M coefficients as a function of frequency of various giant electrostrictors in Fig 3c[26–33]. |Tan δ|$_{EM}$, the electromechanical loss tangent, as function of frequency (Fig 3d) shows a peak at 6-7 kHz for various devices.

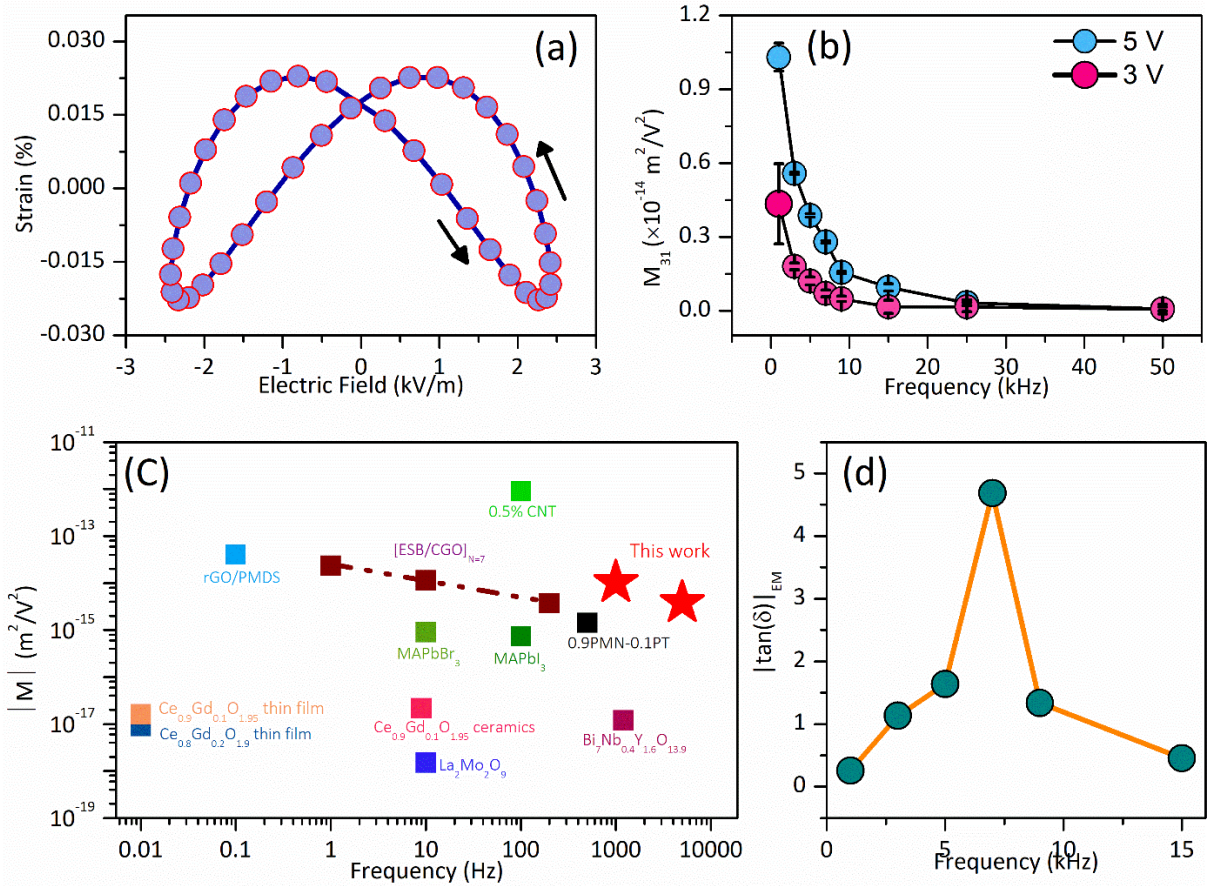

*Fig 3. Analyzing the 2$^{nd}$ order EM response* a) Fourier filtered second harmonic strain response at 5 kHz as a function of varying electric field b) $M_{31}$ coefficients varying with frequency at two different $V_{max}$=3 V (pink) and $V_{max}$=5 V (sky blue) c) Compilation of giant M electrostrictive coefficients previously reported on different material systems, and comparison of the values (indicated as red stars) we achieve in this work, as a function of frequency d) Electromechanical loss tangent (tan δ) at various frequencies, peaking at 6-7 kHz for all the tested devices.

For completeness, let us also note that the electromechanical response measured on vertical metal-insulator-metal capacitors is very weak, owing to the large field drops across the low dielectric constant $TiO_x$ layer (Fig. S3).

**Electrical Characterization**

To understand the correlation of large EM response to dielectric and leakage properties we performed large signal AC I-V measurements from 1 to 50 kHz with voltage varying from -$V_{max}$ to $V_{max}$ (for $V_{max}$=1, 3 and 5 V), and small signal capacitance-voltage-frequency measurements from 1 Hz to 100 kHz. General I-V response of what is referred to as "less leaky" device is shown in Fig 4a (also see Fig S4). Below $V_{max}$=3 V, a good dielectric behavior (capacitive response, phase difference between V and I is close to 90º) is observed, and leakage (resistor) characteristics begin beyond 3 V. A few of the devices tested are leakier than the representative device (leaky I-V shown in Fig S5 a, b), and they showed larger effective $M_{31}$'s (6.15 × 10$^{-14}$ m$^2$/V$^2$) at 1 kHz, 5 V (see Fig S5 c, d). In what follows, we investigate and discuss the "less leaky" devices unless otherwise mentioned. C-*f* data (Fig 4b) shows a relaxation behavior, with the device capacitance reducing significantly with frequency beyond 1 kHz. |Tan δ|$_D$ dielectric loss tangent shows a major peak at ~100 kHz, and a smaller hump at 6-7 kHz frequency, suggesting at least two different RC time constants in the device (Fig S4e). It is important to note that second order EM loss peak correlates with the dielectric loss peak at 6-7 kHz frequency. Reduction in |$M_{31}$| from 1 to 50 kHz also correlates with the decrease in capacitance values.

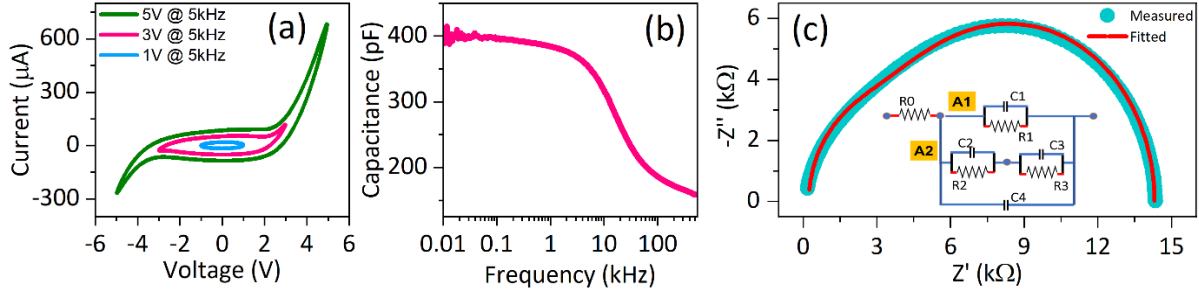

*Fig 4. **Large signal AC I-V characterization and impedance measurements:*** *a) Large signal I-V characteristic on inter digitated electrodes (IDE) at three different voltages ($V_{max}$=1, 3, 5 V). Leakage characteristics begin beyond $V_{max}$=3 V. b) Small signal capacitance measurement as a function of frequency. c) Measured impedance and corresponding fit of the equivalent circuit (inset), detailed circuit is shown in Fig S6a)*

To glean better insights into what dielectric relaxations are correlated to EM response in our complex system, we modeled our impedance spectroscopy data with equivalent circuit (shown in inset of Fig 4c). Our IDE was simplified as lateral device with two terminals, with one terminal ground and the other one sourced (at voltage V). In this configuration, voltages at various nodes are represented in Fig S6a. Our system is modeled as two parallel R||C circuits, referred to as A1 and A2 (see Fig S6a) overall in parallel with interfacial Maxwell-Wagner[34] capacitance. A1 represents lateral BTO layer which exhibits various defect induced relaxation mechanisms (from polar nano-like regions shown in Fig 1c, from interfacial charges at the twin boundaries[12,35] shown in Fig S2a, and electroactive twin boundary motion). A2 contains two different R||C elements in series. The first lumped element represents vertical field drop across BTO (voltage drop from source electrode to BTO-TiO$_x$ interface: V-V$_2$, as well as from BTO-TiO$_x$ interface to the ground: V$_2$-V$_3$ in Fig S6a), while the second element represents voltage drop vertically across TiO$_x$ layer (also see Supplementary note). At frequencies <50 kHz, we obtain very good fits based on such a model as shown in Fig 4c and Fig S4e. Our model suggests that the dielectric relaxations in the BTO layer show a RC time constant of 6-7 kHz, at which frequency we also observed a peak in the |tan δ|$_{EM}$ response of the electromechanical behavior. Thus, the features responsible for electromechanical and dielectric response in BTO correlate very well.

**Discussion:**

The second order EM behavior is an effect of one or more of the following phenomena: a) ferroelectric switching, b) thermal expansion because of device heating, c) non-classical defect-induced electrostriction. Our defective BTO is not ferroelectric, as we do not see any ferroelectric switching peaks in large signal AC I-V plots (Fig 4a and S4 a, b, c). We also measure a large coefficient of thermal expansion (CTE), $\alpha_{33}$=2.36 × $10^{-5}$/K, through in situ XRD measurements on these films (results published elsewhere[24], also reproduced in Fig S6c). Coincidentally, large CTE and large electrostriction are both related to lattice anharmonicity, in this case induced by defects. The large CTE also means that an increase in device temperature by ~15-30 K can already result in ~100 pm expansion or 0.02% strain, observed on our representative devices (Fig 3a). Device temperatures increase owing to leakage-induced Joule heating or dielectric loss, both of which are included in the in-phase component of current with voltage in the large signal AC measurements. To understand the effect of heating, we simulated the device temperature rise using electrothermal modeling via LTspice. The effective electrothermal circuit is shown and described in Fig S6b. The devices were modeled as linear resistors, with resistance (R) given by the ratio of maximum voltage to maximum current. In reality, our devices are non-linear resistors, and this underestimates the resistance, and eventually overestimates the device temperature range. The $V^2$/R power was fed into the thermal circuit as an input heat source, when voltage was cycled at various frequencies from -$V_{max}$ (3 V and 5 V) to $V_{max}$. We show that for a representative "less leaky" device whose I-V characteristics are shown in Fig 4a (also see Fig. S4(a, b, and c) for I-t at various $V_{max}$), the max device temperature rise is 0.8-0.9 K at 1 kHz and 5 V (see Fig S7a), and it reduced with increase in frequency (see Fig S7b). These results clearly show that major contributor to large EM response is electrostriction, and not device heating. We note here that for devices which are leakier (I-V characteristics shown in Fig S5a, b), the max displacements (~500 pm at 5 V, 1 kHz, compare with 150 pm at same conditions for less leaky devices) and effective |$M_{31}$| coefficients observed are larger (Fig S5c, d), and device temperature increases to ~ 10 K (Fig S8a, b), which contributes to only about 60 pm of the displacement out of 500 pm measured. Thus, electrostriction induced by defects is the major contributor to the total electromechanical response in all our devices. We stipulate that the concentration of defects (point defects as well as twin boundaries) in leaky devices is larger, which also correlates with the observation of larger electrostriction.

When 5 V is applied across the IDE, the field ($E_1$) in the BTO layer can be approximated to be 2.5 kV/cm. At such a small field (on non-leaky devices), we obtain 100s of pm vertical displacement, resulting in giant $M_{31}$s. Through the analysis presented so far, the most important contributor to this second order effect is defect-induced electrostriction (and not Joule heating and ferroelectricity). We note that the decrease in $|M_{31}|$ with frequency is correlated with the decrease in device capacitance in 1 to 50 kHz frequency range. Furthermore, dielectric and electromechanical loss tangents also correlate peaking at 6-7 kHz. Our impedance spectroscopy modeling reveals that such a behavior is a consequence of relaxation mechanisms occurring in polar nano-like regions induced relaxations and at the twin boundaries of the BTO layer. Furthermore, the mobility of electroactive twin walls also contributes to the enhanced electromechanical responses as is the case in ferroelectric materials[12]. It may be noted that the fundamental origin of all these relaxations and the correlated giant electrostrictive behavior at these frequencies is due to structural and polarization disorder created by Ba and oxygen non-stoichiometry, and associated lattice anharmonicity (also evidenced by large CTE). The giant $|M_{31}|$ coefficients at larger frequencies are a consequence of large electroactive defect induced polarizabilities, coupled elastic dipoles and long-range coherent strain fields in addition to possible electroactive twin wall motion.

**Conclusions**:

In this work, we demonstrate giant electrostrictive coefficients upto 5 kHz for defective barium titanate films epitaxially integrated with Si. These are record values reported on any materials system beyond frequencies of a few 100s of Hz. We also show that such large response is robust and exists even upon cycling for thousands of times. We show that these coefficients and corresponding EM losses are very much correlated with dielectric relaxations induced by various defects, which fundamentally introduce large lattice anharmonicity. This lets us propose that in order to achieve giant EM responses at even higher frequencies, it is worth to first explore defect-engineering strategies aimed towards reducing the RC time constants of defect-induced dielectric relaxation mechanisms. In addition, we also propose that in addition to point defects, 2D defects that are mobile (just as in ferroelectrics) help in achieving larger electrostrictive responses at higher frequencies. Our work provides a significant step forward in expanding the bandwidth of giant EM responses in Si compatible, lead-free materials.

## Methods:

### Synthesis of epitaxial stack of defective BaTiO$_3$/TiN on Si

**Stage 1:** An epitaxial TiN template on Si is deposited first. 40-60 nm epitaxial TiN on n++ Si (100) was deposited by reactive pulsed laser deposition (ablating a Ti target (99.5% pure, GfE, GmBH, Germany) in N$_2$ (99.9999% pure) ambient in an eclipsed off-axis configuration. The epitaxial growth of TiN on Si (100) and the geometry are discussed in detail in ref[23]. The laser fluence and repetition rate are 1.5 J/cm$^2$ and 20 Hz, respectively. The substrate temperature, target to substrate distance and the chamber pressure during TiN deposition are 700 °C, 3.0 cm and 0.6 mbar, respectively.

**Stage 2:** The 175-245 nm BaTiO$_3$ was deposited on the epitaxial TiN/Si(100) by PLD in a different chamber equipped with reflection high energy electron diffraction (RHEED) (STAIB Instruments, GmBH, Germany, Model: Torr RHEED) operated at 30 kV to monitor the growth surface. Prior to deposition the TiN/Si is dipped in HF:H$_2$O (1:10 V/V) for 30 sec to remove the TiO$_x$ formed on the surface[23]. The laser fluence and repetition rate are 1.5 J/cm$^2$ and 2 Hz, respectively. The substrate temperature and chamber pressure are 600 °C and 5 × 10$^{-6}$ mbar, respectively during the deposition and cooled to room temperature in an oxygen ambient at a pressure of 0.1 mbar. Post deposition the sample is annealed at 500 °C for 1 hour at atmospheric pressure with an oxygen flow rate of 3 slm.

### Structural and composition characterization

**X-ray Diffraction and X-ray Photo-emission spectroscopy**: The crystal structure of BTO/TiN/Si was investigated using 4-circle X-ray diffractometer using a Cu-K$_\alpha$ source (1.5402 Å) (Rigaku Smart Lab). Chemical composition of the BTO films was investigated using XPS (Kratos Axis Ultra) equipped with a monochromatic Al X-ray source. To determine the BTO film stoichiometry, a 10mm circular disc of bulk BTO pellet was used as a reference during the XPS. The survey and high resolution XPS spectra were acquired with 1 eV and 0.1 eV resolution, respectively. The in-situ XPS depth profiling for BTO/TiN/Si samples and BTO pellet was done using an Ar$^+$ ion beam with an energy of 4 kV. C1s peak (284.6 eV) was used to calibrate the survey and high resolution XPS spectra. BTO film composition was calculated from survey spectra and the background data was modeled using Shirley algorithm.

**STEM Imaging and Analysis**: The cross-section FIB lamella for TEM analysis was prepared using a focused ion beam (Model: FEI, Scios2) and is investigated using a double aberration corrected Thermofisher Themis microscope operated at 300 kV, and a non-aberration corrected Themis microscope also operated at 300 kV, equipped with chemi-STEM EDS system. STEM-EDS was performed on non-aberration corrected Themis microscope, and data was acquired until sufficient counts (SNR>5) was obtained from binned pixels. IDPC-STEM images were obtained using a four-segment anuular bright field detector collecting signal from 6-20 mrad. This is a linear imaging technique with contrast ~Z, and thus is sensitive to lighter elemental columns such as oxygen.

**Polarization mapping analysis and oxygen vacancy identification**

Atomic resolution HAADF-STEM images were used to map Ti displacements in every unit cell. To estimate and quantify such displacements a clean scan distortion free lattice image was selected, which were then Bragg filtered for final mapping. Finally, Ti displacement mapping away from the center of mass of Ba (A-site) unit cell was performed using Atomap and Temul-toolkit. Finally, these displacements are represented by arrows (indicating both magnitude and direction of displacement) overlaid on the image. Ti displacements are a good estimator for unit cell dipole moments in $BaTiO_3$.

**Device fabrication**

The patterns for electrical contacts were defined using optical lithography (Heidelberg) and Cr/Au (5/60 nm) contacts were deposited using DC sputtering. For out of plane measurements the contact pad size of $100 \times 100$ μm$^2$ was used. For in-plane measurements interdigitated contacts which consists of 4 pairs of fingers with length of 80 μm, width of 20 μm and spacing between the fingers of 20 μm were used (Fig S2b and c).

**Electromechanical response and I-V characteristics**

Laser Doppler Vibrometer (LDV) (model: MSA 500) equipped with a 532 nm reference and probe laser was used for studying both in plane and out of plane electromechanical response of the $BaTiO_3$ films. Displacement response for different frequencies was obtained for 30-40 msec as a function of time is averaged over several cycles (250-500). This response at each frequency was further averaged to obtain single waveform with standard deviations shown as error bars in M coefficient plots. So all in all, every tested device was cycled for >5000 cycles depending on the frequency of operation to obtain one data set as shown in Fig 3.

Frequency dependent IV and Impedance measurements were performed using MFIA Impedance Analyzer. Nyquist plots were fitted using z-fit utility of EC-Lab Demo. To estimate the rise in device temperature, electro-thermal simulations were carried out using LTspice simulator (Analog Devices).

## Acknowledgements

This work was partly carried out at Micro and Nano Characterization Facility (MNCF), and National Nanofabrication Center (NNfC) located at CeNSE, IISc Bengaluru, funded by NPMAS-DRDO and MCIT, MeitY, Government of India; and benefitted from all the help and support from the staff. P.N. acknowledges Start-up grant from IISc, Infosys Young Researcher award, and DST-starting research grant SRG/2021/000285. The authors acknowledge funding support from the Ministry of Human Resource Development (MHRD) through NIEIN project, from Ministry of Electronics and Information Technology (MeitY) and Department of Science and Technology (DST) through NNetRA and the Thematic Unit of Excellence for Nano Science and Technology project from DST Nano Mission. PN would like to acknowledge discussions with Evgenios Stylianidis from University College London. All the authors acknowledge the usage of national nanofabrication center, micro nano characterization center, and advanced facility for microscopy and microanalysis of IISc for various fabrication and characterization studies.

## Author contributions

Ideation began with discussions between PN, SV, SP. They designed the experiments. SV synthesized the samples. SKP, SP, SV, UK, SHM carried out the electromechanical measurements. SKP, RKR, MA, AM, SV carried out STEM analysis, with data analysis routines run by RKR and SKP. SV carried out XRD, XPS and corresponding data analysis. UK, SKP, RSS carried out impedance spectroscopy, impedance device modeling and electrothermal modeling using SPICE with supervision from SC and PN. GP, MJ performed complimentary COMSOL electrothermal simulations and independently confirmed the SPICE results. VK, SV fabricated the devices. BK, PN, MA supervised STEM analysis; SA, SC, PN supervised the electrical and electromechanical measurements and corresponding analysis; SR, PN supervised the growth and characterization aspects. PN managed the coordination between the various teams. SKP, PN, SV, SP cowrote the manuscript with inputs from UK. All the authors read and commented on the manuscript.

**Supplementary Information**

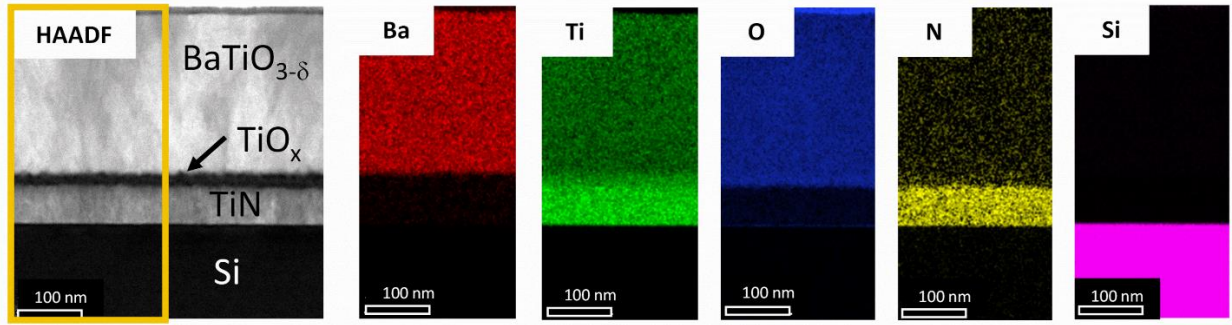

**Fig S1 Chemical maps of various layers in the heterostructure:** EDS map showing elemental distribution (intensities correspond to atom%) in BTO, TiN, layers, and Silicon substrate.

**Note 1:** Fig S1 shows the EDS map acquired on the cross-section FIB lamella of sample stack Barium Titanate (BTO) grown on Titanium Nitride (TiN) buffered Si (100). It can be seen that an additional $TiO_x$ interfacial layer of thickness ~20 nm is formed between TiN and BTO.

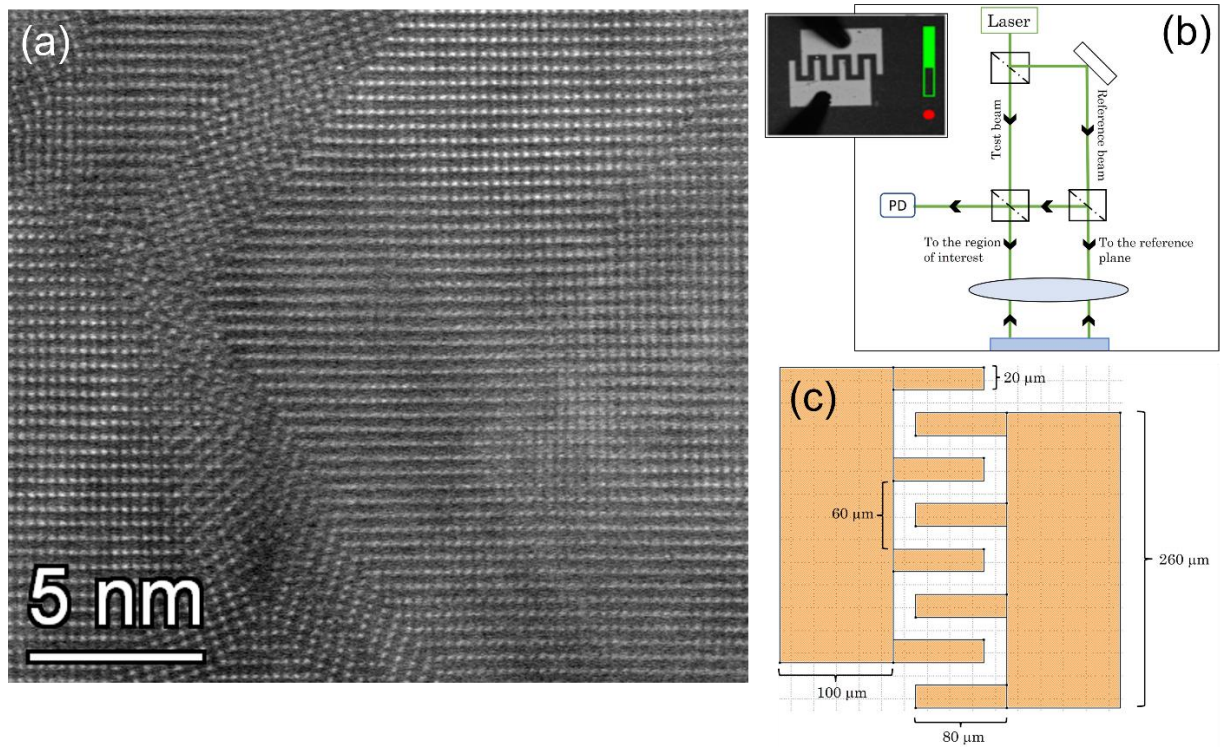

**Fig S2: Defects, electromechanical measurement scheme and device structure** (a) HAADF-STEM image along [110] zone axis containing domain boundaries which could be one of the contributors to MW type of relaxation mechanisms coming from the BTO layer. (b) Laser doppler vibrometer setup and optical image (inset) of Interdigitated Electrodes and (c) corresponding detailed feature dimensions of IDE.

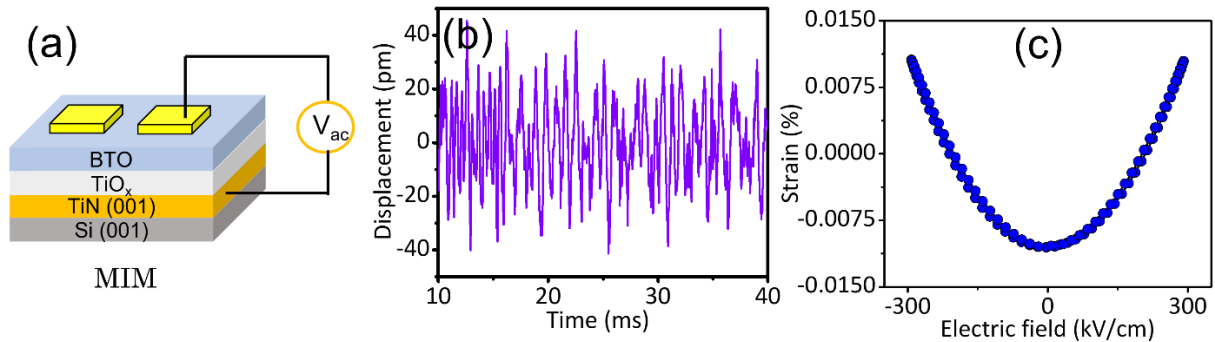

**Fig S3: Electromechanical response on vertical capacitors.** (a) Metal insulator metal (MIM) capacitor schematic (b) Vertical displacement as a function of time of MIM capacitor with BTO thickness 170 nm and (c) second harmonic response with a varying electric field (5 V) at 1 kHz.

**Note 2:** The strain response when electric field is applied in the vertical direction (on MIM vertical capacitors) is much smaller (18 pm displacement at 0.30 MV/cm field on the device) as compared to the response recorded with lateral electric field on IDE devices. This is a result of very small vertical field drop across the BTO layer and a large drop across low dielectric constant TiOx layer. As a result, in this geometry the accurate values of $M_{33}$ were not estimated and we concentrate rather on $M_{31}$

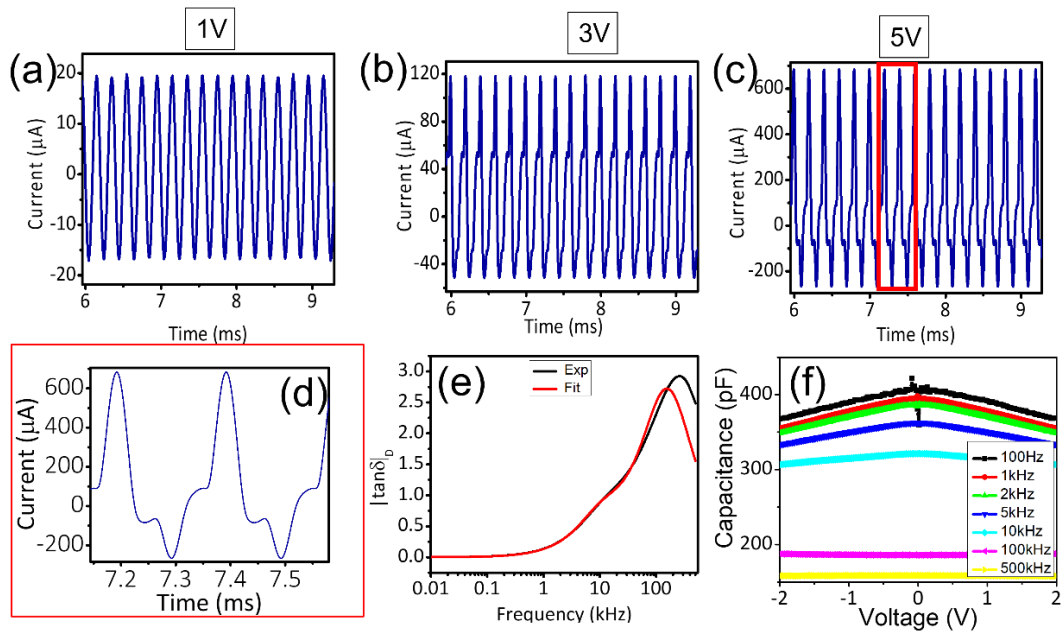

**Fig S4: Large signal I-V and small signal C-V measurements.** Current response as a function of time for three different voltages (a) at 1 V (b) 3 V, and (c) 5 V. (d) Magnified region from marked red enclosure in Fig (c) showing non-linear current response. (e) Dielectric loss tangent tan δ and corresponding fit obtained for equivalent circuit (f) CV measurements as a function of frequency. Notably a peak at 0V in both forward and reverse sweeps upto 10 kHz indicates non ferroelectric, yet very soft dielectric response.

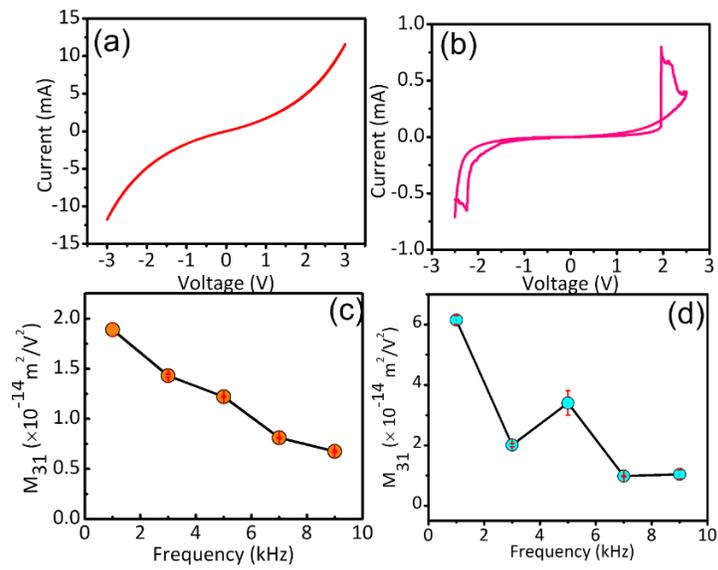

**Fig S5: DC I-V characteristics of leaky devices and their electromechanical response** I-V characteristics (a) and (b) from "leakier" devices showing even larger $M_{31}$ coefficients (c) and (d) respectively which are two and six times than the $M_{31}$ coefficients for the "less leaky" device at 1 kHz frequency

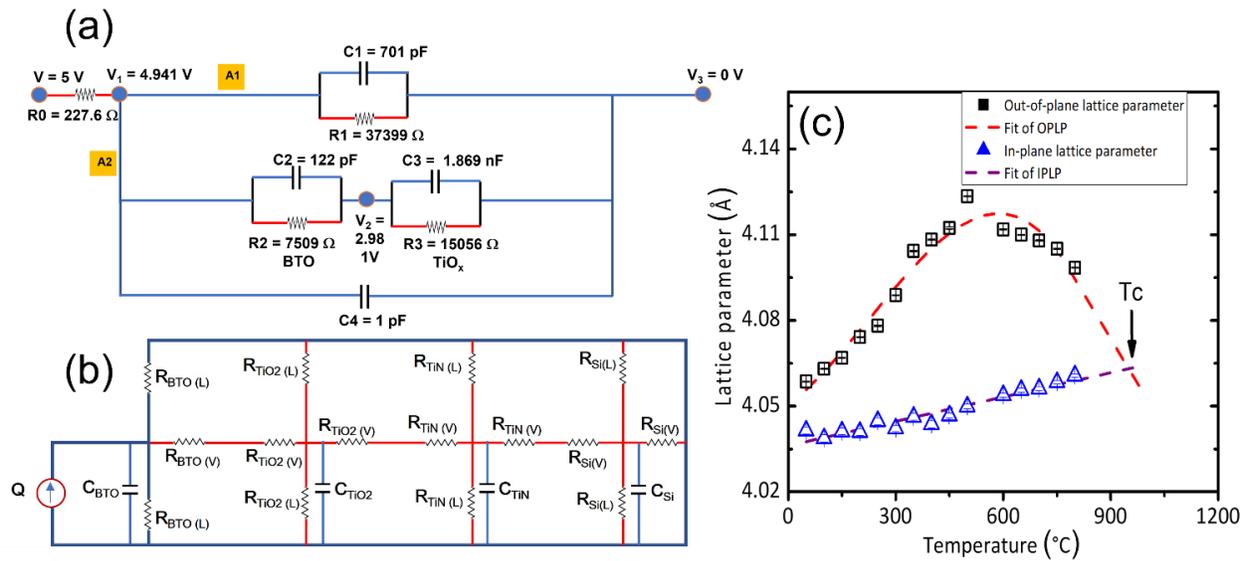

**Fig S6: Impedance and thermal circuit models** (a) Equivalent circuit used for fitting the impedance response including values of RC elements corresponding to lateral BTO layer (top path or A1) and A2 (bottom path) that includes vertical drop across BTO ($V_1$-$V_2$), TiO$_x$($V_2$-$V_3$) and the MW contribution from interface obtained from Z-fit. The RC time constant of lateral BTO path is 6-7 kHz and correlates with the peak in both dielectric and electromechanical tan δ. (b) Equivalent thermal circuit model used for electrothermal simulations performed on LTspice. (L) and (V) represents lateral and vertical element respectively in each layer (c) In-situ XRD of the defective BTO/TiN/Si stack carried out to determine the coefficient of thermal expansion (reproduced[1])

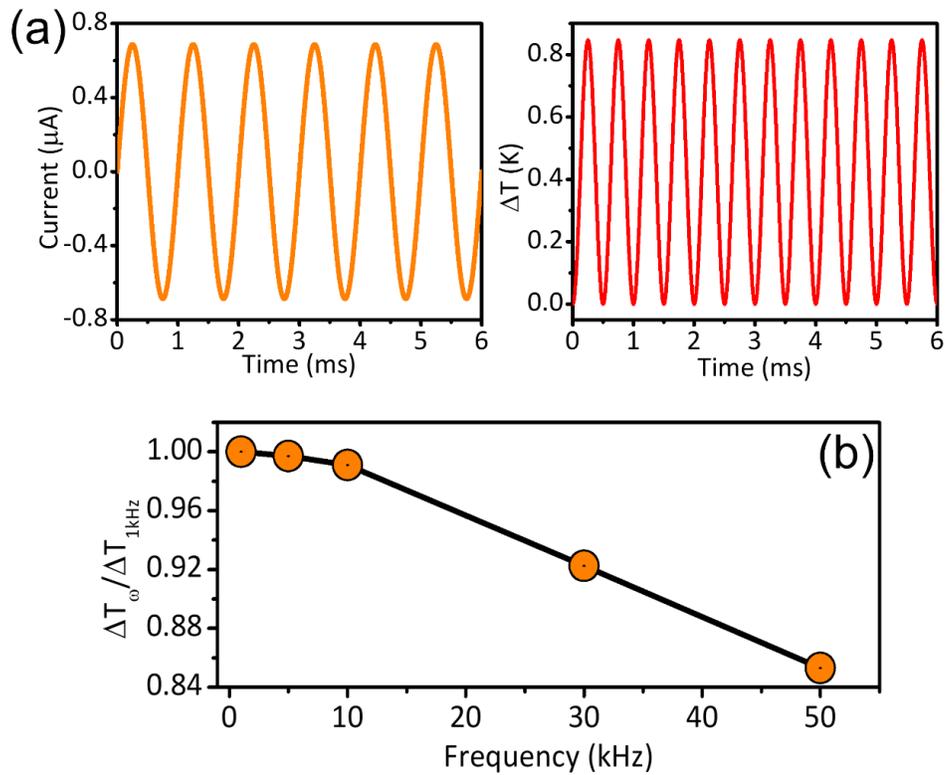

**Fig S7. Results of device thermal simulations on less-leaky devices:** Current waveform (from I-V measurement) used for simulating the temperature change in "less leaky" device as a function of time shown for (a) 1 kHz 5V (b) normalized temperature change as a function of frequency (ω). Maximum temperature rise is 0.8 K suggesting that Joule heating is not important in these less leaky devices for explaining the large electromechanical effect.

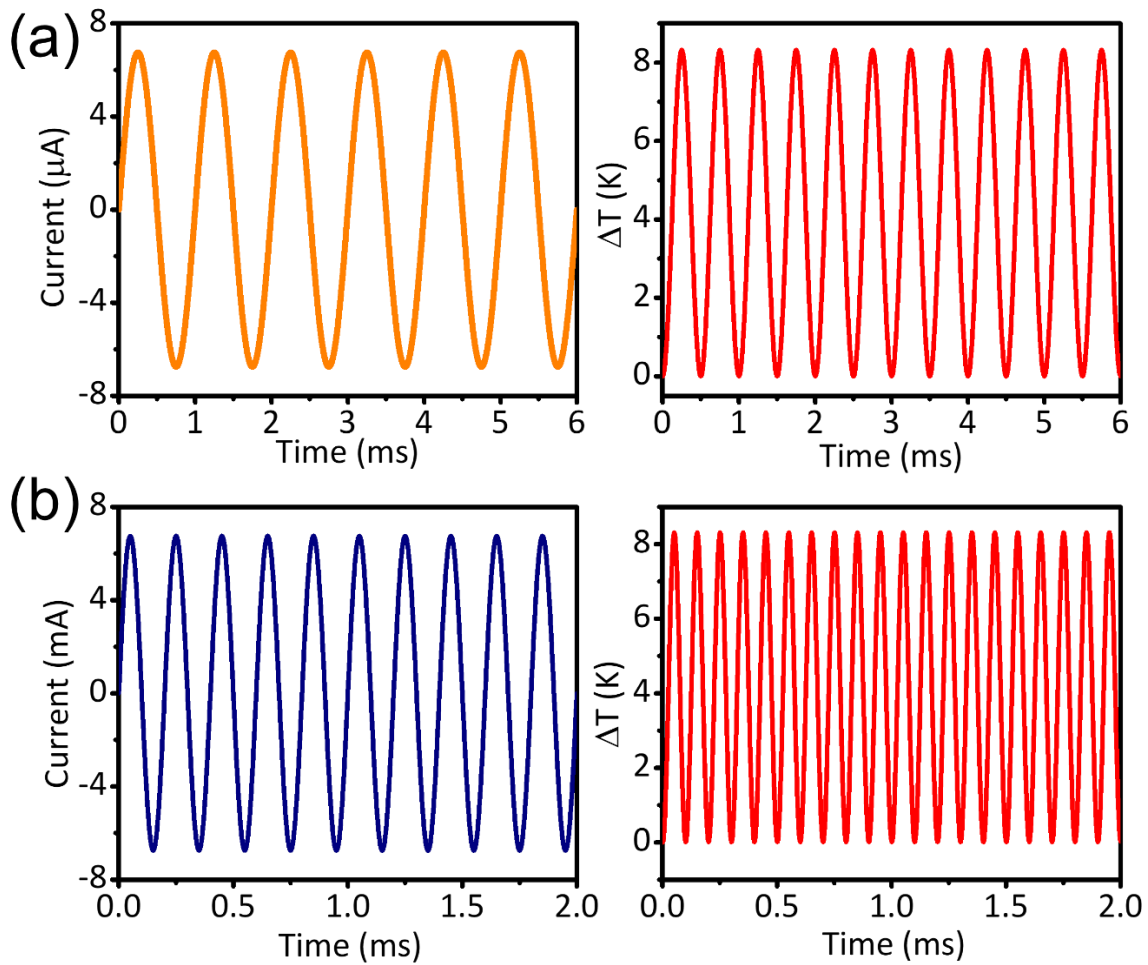

**Fig S8. Thermal simulations on leaky devices:** Current waveform (from *I*-V measurement) used for simulating the temperature rise in "leakier" device as a function of time shown for (a) 1kHz, 5V (b) 5kHz, 5V. Here we observe a maximum temperature rise of 8 K, which corresponds to a film displacement amplitude of ~60 pm. Note that the total displacement amplitude measured in this device at 1 kHz, 5 V is ~500 pm. Joule heating is still a minor part in explaining the large electromechanical effects.